\documentclass[12pt,a4paper,english,amssymb,nofootinbib,superscriptaddress]{revtex4}
\usepackage{lmodern}

\usepackage[T1]{fontenc}
\usepackage[latin9]{inputenc}
\usepackage{slashed}
\usepackage{babel}
\usepackage{amsmath}
\usepackage{slashed}
\usepackage{amssymb}
\usepackage{graphicx}
\usepackage{epstopdf}
\usepackage{subfigure}
\usepackage{esint}
\usepackage[unicode=true,pdfusetitle,
bookmarks=true,bookmarksnumbered=false,bookmarksopen=false,
breaklinks=false,pdfborder={0 0 1},backref=false,colorlinks=false]
{hyperref}

\makeatletter
\@ifundefined{textcolor}{}
{%
	\definecolor{BLACK}{gray}{0}
	\definecolor{WHITE}{gray}{1}
	\definecolor{RED}{rgb}{1,0,0}
	\definecolor{GREEN}{rgb}{0,1,0}
	\definecolor{BLUE}{rgb}{0,0,1}
	\definecolor{CYAN}{cmyk}{1,0,0,0}
	\definecolor{MAGENTA}{cmyk}{0,1,0,0}
	\definecolor{YELLOW}{cmyk}{0,0,1,0} 
}
 
\usepackage{latexsym}\usepackage{bm}

\makeatother

\makeatother 

\begin{document}
\title{ On memory effect in modified gravity theories}
\author{Ercan Kilicarslan $^{1,*}$\\$^{1}$Department of Physics,\\
 Usak University, 64200, Usak, Turkey \\
$^*$Correspondence:ercan.kilicarslan@usak.edu.tr }

\maketitle

\noindent{\textbf{Abstract:}} In this note, we discuss the gravitational memory effect in higher derivative and infinite derivative gravity theories and give the detailed relevant calculations whose results were given in our recent works. We show that the memory effect in higher derivative gravity takes the same form as in pure GR at large distances, whereas at small distances, the results are different. We also demonstrated that, in infinite derivative gravity, the memory is reduced via  error function as compared to Einstein's gravity. For the lower bound on the mass scale of non-locality, the memory is essentially reproduces the usual GR result at distances above at very small distances. 

\noindent{\textbf{1. Introduction}}

 \noindent Reconciling unitarity (i.e., ghost and tachyon freedom) with renormalizability in gravity theories has been foremost obstacle to obtain a complete theory of gravity. By adding scalar quadratic curvature terms to Einstein's theory such as the higher derivative gravity, $ R+\alpha R^2+\beta R_{\mu\nu}^2$, renormalizability is restored, but theory does not satisfy the requirements of the unitarity due to a contradiction between the massless and massive spin-$2$ modes \cite{stelle} (For the detailed discussion, see \cite{Odintsov}). Thus, the theory has spin-2 Weyl ghost mode that leads to Ostragradsky type instabilities at the classical level which become ghosts at the quantum level. As a consequence, the addition of higher order curvature terms give rise to a contradiction between the unitarity and the renormalizability. 
On the other hand, another vigorous attempt has recently been proposed as a ghost and singularity-free theory of gravity. This theory, is called infinite derivative gravity (IDG), has the potential to provide a viable theory \cite{Biswas1,Biswas2}. Here the action is built from non-local  analytic functions $F_i(\Box)$ [given in Eq.(\ref{idfunc})], where $\Box$ is the d'Alembartian operator ($\Box=g^{\mu\nu}\nabla_\mu\nabla_\nu$)\footnote{For recent progresses on IDG, see \cite{Tomboulis,Biswas4, Modesto,Biswas5,Biswas6,Buoninfante:2018xiw,Modesto1,Modesto2,Modesto22,Modesto3, Talaganis:2016ovm,Edholm:2016hbt, Kilicarslan:2018yxd, Modesto4,Boos:2018bxf,Boos:2018kir,Moffat1,Moffat2,Briscese:2012ys, Giacchini:2018wlf}.}. In IDG, the propagator in a Minkowski background is given as 
\begin{equation}
\Pi_{IDG}=\frac{P^2}{a(k^2)}-\frac{P_s^0}{2a(k^2)}=\frac{\Pi_{GR}}{a(k^2)},
\end{equation}
in which $P^2$ and $P_s^0$ are Barnes-Rivers spin projection operators \cite{Biswas1} and $\Pi_{GR}$ is the graviton propagator in pure GR. Also, arbitrary function $a$ is given in terms of $F_i(\Box)$ [see Eq.(\ref{relations})]. In order to the theory to be ghost-free and have not extra scalar dynamical degrees of freedom (DOF) other than the massless graviton propagating in $3+1$ dimensions, $a(k^2)$ term should have no roots. For this purpose, $a(k^2)$ can be chosen to be an exponential of an entire function as $a(k^2)=e^{\gamma(\frac{k^2}{M^2})}$, where $\gamma(\frac{k^2}{M^2})$ is an entire function. This choice provides that the propagator has no new extra dynamical roots compared to pure GR and thus it is only modified by an arbitrary function $a(k^2)$. In the $a(k^2)\to 0$ or $k\ll M$ limit, the propagator reproduces the usual GR result. Moreover, the theory is free from Ostragradsky's ghost instabilities since the modified propagator does not contain any extra dynamical DOF. With the modified propagator, the infinite derivative extension of GR has upgraded small scale behavior. For example, it has been recently studied in \cite{Edholm:2016hbt} that IDG has non-singular Newtonian potential for point source as one approaches $r\to 0$.  In \cite{Kilicarslan:2018yxd}, the discussions is extended to the case where there are spin-spin and spin-orbit interactions in addition to mass-mass interactions, it is shown that not only mass-mass interaction but also spin-spin and spin orbit interactions are regular and finite at the origin. Hence, the theory is very well-behaved in the small scale unlike GR. On the  other hand, loop divergences beyond $1$-loop for IDG would be regulated by introducing some appropriate form factors \cite{Talaganis}. Additionally, IDG has also a potential to solve the problem of singularities in black holes and cosmology \cite{Biswas1, Biswas2,Tomboulis,Biswas4, Modesto,Biswas5,Biswas6,Buoninfante:2018xiw}.  

In this work, we would like to discuss the gravitational memory effect in higher derivative gravity and IDG in a flat spacetime and compare these with the result of GR. At this point, one can ask that what is memory effect. Let's give the brief summary: gravitational waves, created by merger of neutron stars or black holes etc, induce a non-trivial effect on a system composed of inertial test particles. In other words, a pulse of gravitational wave produce a non-trivial change in the relative separation of test particles. This phenomenon is known as  the gravitational memory effect and comes in two forms: ordinary (or linear) \cite{zeldovich} and null  (or
non-linear) \cite{Christodoulou}. Recently, many works have been done on memory effect in various aspects \cite{Garfinkle,Satishchandran:2017pek,Pasterski,Strominger1,Strominger2,Flanagan,Hollands,Tolish1,Tolish2,Bieri2,Zhang,Zhang2,Kilicarslan}. In fact, gravitational memory effect was given as a result in higher derivative gravity \cite{Kilicarslan} and IDG \cite{Kilicarslan:2018yxd} since the calculations are tedious and lengthy. In this paper, we shall go further and give detailed relevant computations of gravitational memory effect in these theories. 

The layout of the paper is as follows: In Sec. 2, we calculate the memory effect in higher derivative gravity and investigate the effects of quadratic terms on the memory. Section 3 is devoted to compute the memory effect in IDG and its large and small distance limits. In that section, we also consider the effects of mass scale of non-locality on gravitational memory. 

 \noindent{\textbf{2. Memory effect in higher derivative gravity}}

 \noindent In this section, we will study the detailed computations for the memory effect in generic even dimensional flat backgrounds. To do so, let us first notice that the action of higher derivative gravity is given as follows 
\begin{eqnarray}
I & = & \int d^{{D}}x\,\sqrt{-g}\left\{
\frac{1}{\kappa}R+\alpha R^{2}+\beta
R_{ab}^{^{2}}
+\gamma\left(R_{abcd}^{2}-4R_{ab}^{2}+R^{2}\right)+ {\cal {L}}_{\mbox{matter}}\right\} ,
\label{action}\end{eqnarray}
where $\kappa$ is Newton's constant\footnote{In this part, for the sake of simplicity, we will use the abstract index notation\cite{R.M.Wald} and geometric unit system ($G=1$).}. The source coupled field equations reads 
\begin{eqnarray}
\frac{1}{\kappa}\left(R_{ab}-\frac{1}{2}g_{ab}R  \right )
+2\alpha R\left(R_{ab} -\frac{1}{4}g_{ab}R\right)+\left(2\alpha+\beta\right)\left(g_{ab}\square-\nabla_{a}\nabla_{b}\right)R\nonumber\\
+2\gamma\left[RR_{ab}-2R_{acbd}R^{cd}+R_{acde}R_{b}^{\;\;cde}
-2R_{ac}R_{b}^{\;\;c}-\frac{1}{4}g_{ab}\left(R_{cdef}^{2}-4R_{cd}^{2}
+R^{2}\right)\right]\nonumber\\+\beta\square\left(R_{ab}-
\frac{1}{2}g_{ab}R\right)+2\beta\left(R_{acbd}
-\frac{1}{4}g_{ab}R_{cd}\right)R^{cd} ) =\tau_{ab}.
\label{fieldequations}
\end{eqnarray}
Linearization of the field equations (\ref{fieldequations}) about the Minkowski background metric, $g_{\mu\nu}=\bar{g}_{\mu\nu}+h_{\mu\nu}$, yields \footnote{We will work with the mostly plus signature $\eta_{\mu\nu}=\mbox{diag}(-1,1,1,1)$.} \cite{Gullu-Tekin}
\begin{eqnarray}
T_{ab}\left(h\right) & = & \frac{1}{\kappa} {\mathcal{G}}_{ab}^{L}
  +  \left(2\alpha+\beta\right)\left(\bar{g}_{ab}\bar{\square}-\bar{\nabla}_{a}\bar{\nabla}_{b}
 \right)R^{L} +  \beta\bar{\square}{\mathcal{G}}_{ab}^{L}
,\label{linearizedfirst}
\end{eqnarray}
in which $T_{ab}\left(h\right)$ is conserved energy momentum tensor which includes all the quadratic order terms as $ T_{ab}=\tau_{ab}+\Theta(h^2,h^3,...) $, L refers to linearization and  ${\mathcal{G}}_{ab}^{L}$ is the linearized Einstein tensor:
\begin{equation}
{\mathcal{G}}_{ab}^{L}=R_{ab}^{L}-\frac{1}{2}\bar{g}_{ab}R^{L}.
\label{einstein}\end{equation}
Here the linearized Ricci tensor $R_{ab}^{L}$ and the scalar curvature $R^{L}$ are given respectively as \cite{deser_tekin}
\begin{equation}
R_{ab}^{L}=\frac{1}{2}\left(\bar{\nabla}^{c}\bar{\nabla}_{a}h_{bc}
+\bar{\nabla}^{c}\bar{\nabla}_{b}h_{ac}-\bar{\square}h_{ab}-\bar{\nabla}_{a}\bar{\nabla}_{b}h\right), \,\,
R^{L}=-\bar{\square}h+\bar{\nabla}^{a}\bar{\nabla}^{b}h_{ab}.
\label{lineartensor}\end{equation}
By using the linearized form of the tensors and then manipulation of (\ref{linearizedfirst}) reads
\begin{equation}
\left [ \left( 4 \alpha (D-1) + D\beta \right)\bar{\square} - (D-2)\left (\frac{1}{\kappa}  \right ) \right ] R^L 
 = 2 T.
\label{manipedenk}
\end{equation}
In the de-Donder gauge $\partial^{a}h_{ab}=\frac{1}{2} \partial_{b}h $ which give rises to $ R^L= -\frac{1}{ 2 }\partial^2h_{ab} $ and $ G_{ab}^{L}= -\frac{1}{ 2 }\partial^2(h_{ab}-\frac{1}{2} \bar{g}_{ab}h)$. By using of these, the field equations take the form
\begin{equation}
 (\frac{1}{\kappa}+ \beta \partial^2) \partial^2  h_{ab}= -2T_{ab}+ 2(2\alpha+\beta)(\bar{g}_{ab}\partial^2-\partial_a \partial_b)R^L
 - (\frac{1}{\kappa}+\beta \partial^2) \bar{g}_{ab}R^L,
\end{equation}
which is the equation that we shall work with for this section. Note that by using Eq.(\ref{manipedenk}), this equation can be recast as in the following desired form
\begin{equation}
\begin{aligned}
 h_{a b}=& -\frac{2T_{ab}}{( \beta\partial^2+\frac{1}{\kappa} ) \partial^2 } +\frac{4(2\alpha+\beta)}{( \beta\partial^2+\frac{1}{\kappa} )\bigg(\left( 4 \alpha (D-1) + D\beta \right)\partial^2-\frac{1}{\kappa}(D-2) \bigg) \partial^2}(\bar{g}_{ab}\partial^2-\partial_a \partial_b)T\\&-\frac{2\bar{g}_{ab}T}{\bigg(\left( 4 \alpha (D-1) + D\beta \right)\partial^2-\frac{1}{\kappa}(D-2) \bigg) \partial^2} ,
 \end{aligned}
\end{equation}
whose retarded inhomogeneous solution can be found to be
\begin{equation}
h_{ab} = \int \bigg(2 G^1(x,x')T_{ab}(x')-4(2\alpha+\beta)G^2(x,x')(\bar{g}_{ab}\partial^2-\partial_a \partial_b)T(x')+2 \bar{g}_{ab}G^3(x,x')T(x')\bigg)d^{D}x',
\end{equation}
where the scalar Green's function is defined as 
\begin{equation}
\begin{aligned}
&G^1(x,x')=\frac{1}{\beta}\bigg(( \partial^2-m_\beta^2) \partial^2 \bigg)^{-1},\\& G^2(x,x')=\frac{1}{\beta\left( 4 \alpha (D-1) + D\beta \right)}\bigg((\partial^2-m_\beta^2) (\partial^2 - m_c^2)\partial^2\bigg)^{-1},\\& G^3(x,x')=\frac{1}{\left( 4 \alpha (D-1) + D\beta \right)}\bigg((\partial^2 - m_c^2)\partial^2\bigg)^{-1},
\end{aligned}
\end{equation}
where $m_\beta$ is the mass of massive spin-$2$ graviton given as $m_\beta^2=-\frac{1}{\beta\kappa}$ and $m_c$ is the mass of massive spin-$0$ graviton defined as $m_c^2=\frac{D-2}{\kappa\left( 4 \alpha (D-1) + D\beta \right)}$.
 To calculate the memory effect, we follow the method of \cite{Satishchandran:2017pek, Garfinkle}. For this purpose, let us now consider the incoming massive particles that interact at the point $t=0, \vec{x}$ and some outgoing massive particles created at this point. Then, the corresponding energy momentum tensor of the particle sources can be written 
\begin{equation}
T_{ab}=\sum_{(j)in}m^{\rm in}_{(j)}u_{(j)a}u_{(j)b}\frac{d\tau_{(j)}}{dt}\delta_{3}(\mathbf{x}-\mathbf{y}_{(j)}(t))\Theta(-t) + \sum_{(i)out}m^{\rm out}_{(i)}\frac{d\tau_{(i)}}{dt}u_{(i)a}u_{(i)b}\delta_{3}(\mathbf{x}-\mathbf{y}_{(i)}(t))\Theta(t),
\label{energy-momentum}
\end{equation}
where $\Theta$ is the step function, $u_{(i)a}$ and $u_{(j)a}$ are normalized four velocities. Coherently, one can show that the propagators can be explicitly described as
\begin{equation}
\begin{aligned}
&G^1(x,x')=\frac{\kappa \delta(t-t'-r)}{2(2\pi)^{\frac{D-2}{2}}}\bigg(\frac{\Theta(t-t')}{r}(-\frac{1}{r}\frac{\partial}{\partial r})^{\frac{D-2}{2}}-\sqrt{\frac{2}{\pi}}(\frac{m_\beta}{ r})^{\frac{D-3}{2}}K_{\frac{D-3}{2}}(m_\beta r)\bigg)\\
& G^2(x,x')=\frac{1}{\beta\left( 4 \alpha (D-1) + D\beta \right)}\bigg[\frac{1}{2(2\pi)^{\frac{D-2}{2}}m_\beta^2m_c^2}\Theta(t-t')(-\frac{1}{r}\frac{\partial}{\partial r})^{\frac{D-2}{2}}\frac{\delta(t-t'-r)}{r}\\&+\frac{\delta(t-t'-r)}{(2\pi)^{\frac{D-1}{2}}(m_\beta^2-m_c^2)}\bigg(\frac{1}{m_\beta^2}(\frac{m_\beta}{ r})^{\frac{D-3}{2}}K_{\frac{D-3}{2}}(m_\beta r)
-\frac{1}{m_c^2}(\frac{m_c}{ r})^{\frac{D-3}{2}}K_{\frac{D-3}{2}}(m_c r)\bigg)\bigg]\\
& G^3(x,x')=-\frac{\kappa}{2(2\pi)^{\frac{D-2}{2}}(D-2)}\bigg(\frac{\Theta(t-t')}{r}(-\frac{1}{r}\frac{\partial}{\partial r})^{\frac{D-2}{2}}-\sqrt{\frac{2}{\pi}}(\frac{m_c}{ r})^{\frac{D-3}{2}}K_{\frac{D-3}{2}}(m_c r)\bigg),
\end{aligned}
\end{equation}
where $K_{\frac{D-3}{2}}(r)$ is the modified Bessel function of the second kind. With these tools, the retarded solution of higher derivative gravity can be obtained up to leading order $\frac{1}{r}$
\begin{equation}
\begin{aligned}
  h_{ab}(x)=&\frac{\kappa}{(2\pi r)^{\frac{D-2}{2}}}\bigg((\frac{\partial}{\partial U})^{\frac{D-4}{2}}-(m_\beta)^{\frac{D-4}{2}}e^{-m_\beta r}\bigg)\bigg(\alpha_{ab}\Theta(U)+\beta_{ab}\Theta(-U)\bigg)\\&+\frac{\kappa\bar{g}_{ab} \bar{g}_{cd}}{(2\pi r)^{\frac{D-2}{2}}(D-2)}\bigg(-(\frac{\partial}{\partial U})^{\frac{D-4}{2}}+(m_c)^{\frac{D-4}{2}}e^{-m_c r}\bigg)\bigg(\alpha^{cd}\Theta(U)+\beta^{cd}\Theta(-U)\bigg)\\&+\frac{2(2\alpha+\beta)\bar{g}_{cd}}{\beta\left( 4 \alpha (D-1) + D\beta \right)(m_\beta^2-m_c^2)}\frac{1}{(2\pi r)^{\frac{D-2}{2}}}\bigg\{-m_\beta^{\frac{D-8}{2}}e^{-m_\beta r}\bar{g}_{ab}\bigg(m_\beta^2(\alpha^{cd}\Theta(U)\\&-\beta^{cd}\Theta(-U))+2m_\beta\delta(U)(\alpha^{cd}-\beta^{cd})\bigg)+(m_c )^{\frac{D-8}{2}}e^{-m_c r}\bar{g}_{ab}\bigg((m_c^2(\alpha^{cd}\Theta(U)\\&-\beta^{cd}\Theta(-U))+2m_c\delta(U)(\alpha^{cd}-\beta^{cd})\bigg)+\frac{(m_\beta^2-m_c^2)}{m_\beta^2m_c^2}K_aK_b(\alpha^{cd}-\beta^{cd})(\frac{\partial}{\partial U})^{\frac{D-4}{2}}\delta(U)\\&+m_\beta^{\frac{D-8}{2}}e^{-m_\beta r}\bigg(m_\beta^2r_ar_b(\alpha^{cd}\Theta(U)-\beta^{cd}\Theta(-U))+m_\beta(\alpha^{cd}-\beta^{cd})(K_ar_b+K_br_a)\delta(U)\\&+(\alpha^{cd}-\beta^{cd})K_aK_b\delta(U)'\bigg)-  m_c^{\frac{D-8}{2}}e^{-m_c r}\bigg(m_c^2r_ar_b(\alpha^{cd}\Theta(U)-\beta^{cd}\Theta(-U))\\&+m_c(\alpha^{cd}-\beta^{cd})(K_ar_b+K_br_a)\delta(U)+(\alpha^{cd}-\beta^{cd})K_aK_b\delta(U)'\bigg)\bigg\}.
  \label{Field1}
  \end{aligned}
 \end{equation}
Here $U  \equiv t -r$ is the retarded time, $K^a \equiv -\partial^aU=t^a+r^a$ and $t^a$ and $r^a=\partial^ar$ are unit vectors. In this setting, we defined 
\begin{equation}
\begin{aligned}
&\alpha_{ab}(\hat{\mathbf{r}}) \equiv\sum_{(i)out}\frac{d\tau^{(i)}}{dt}\Big(\frac{m^{\rm out}_{(i)}}{1-\hat{\mathbf{r}}\cdot\mathbf{v}^{(i)}}\Big)\bigg(u_{a}^{(i)}u_{b}^{(i)}\bigg),\\&
\beta_{ab}(\hat{\mathbf{r}}) \equiv\sum_{(j)in}\frac{d\tau^{(i)}}{dt}\Big(\frac{m^{\rm out}_{(i)}}{1-\hat{\mathbf{r}}\cdot\mathbf{v}^{(i)}}\Big)\bigg(u_{a}^{(i)}u_{b}^{(i)}\bigg).
\end{aligned}
\end{equation}
Since the memory is related to curvature tensor by means of the geodesic equation, one needs to first compute
the linearized Riemann tensor which is defined as
\begin{equation}
R_{abcd}= \partial_{a}\partial_{[d}h_{b]c} - \partial_{c}\partial_{[d}h_{b]a}.
\end{equation}
Finally, to leading order, the linearized Riemann tensor of metric perturbation yields
\begin{equation}
\begin{aligned}
R_{abcd}=&\frac{\kappa}{(2\pi r)^{\frac{D-2}{2}}}K_{[a}\Delta_{b][c}K_{d]}\frac{d^{\frac{D-2}{2}}}{dU^{\frac{D-2}{2}}}\delta(U)-\frac{\kappa}{(2\pi r)^{\frac{D-2}{2}}}(m_\beta)^{\frac{D-2}{2}}\bigg(K_{[a}\Delta_{b][c}K_{d]}\frac{d^2\Theta(U)}{dU^2}\\&+m_\beta K_{[a}\Delta_{b][c}r_{d]}\frac{d\Theta(U)}{dU}+m_\beta K_{[d}\Delta_{b][c}r_{a]}\frac{d\Theta(U)}{dU}+2m_\beta^2r_{[a}\bar{\alpha}_{b][c}r_{d]}\Theta(U)\\&+2m_\beta^2r_{[a}\bar{\beta}_{b][c}r_{d]}\Theta(-U)\bigg)e^{-m_\beta r},
 \label{Riemannhdg}
 \end{aligned}
\end{equation}
where  we defined
\begin{equation}
\begin{aligned}
&\Delta_{ab} = 2\sum_{(i)out} \frac{d\tau_{(i)}}{dt}\Big(\frac{m^{\rm out}_{(i)}}{1-\hat{\mathbf{r}}\cdot\mathbf{v}_{(i)}}\Big)\bigg(q_{ac}u^{c}_{(i)}q_{bd}u^{d}_{(i)}-\frac{q_{cd}u^{c}_{(i)}u^{d}_{(i)}}{D-2}q_{ab}\bigg) \\
& -2\sum_{(j)in}\frac{d\tau_{(j)}}{dt}\Big(\frac{m^{\rm in}_{(j)}}{1-\hat{\mathbf{r}}\cdot\mathbf{v}_{(j)}}\Big)\bigg(q_{ac}u^{c}_{(j)}q_{bd}u^{d}_{(j)}-\frac{q_{cd}u^{c}_{(j)}u^{d}_{(j)}}{D-2}q_{ab}\bigg) ,
\\&
\bar{\alpha}_{ab} = 2\sum_{(i)out} \frac{d\tau_{(i)}}{dt}\Big(\frac{m^{\rm out}_{(i)}}{1-\hat{\mathbf{r}}\cdot\mathbf{v}_{(i)}}\Big)\bigg(q_{ac}u^{c}_{(i)}q_{bd}u^{d}_{(i)}-\frac{q_{cd}u^{c}_{(i)}u^{d}_{(i)}}{D-2}q_{ab}\bigg),\\
& \bar{\beta}_{ab} = 2\sum_{(j)in}\frac{d\tau_{(j)}}{dt}\Big(\frac{m^{\rm in}_{(j)}}{1-\hat{\mathbf{r}}\cdot\mathbf{v}_{(j)}}\Big)\bigg(q_{ac}u^{c}_{(j)}q_{bd}u^{d}_{(j)}-\frac{q_{cd}u^{c}_{(j)}u^{d}_{(j)}}{D-2}q_{ab}\bigg), 
\label{memorytensor}
\end{aligned}
\end{equation}
and $q_{ab}$ is the projector that projects the metric onto $S^{D-2}$. The relative separation between two massive test particles at rest is given by geodesic deviation equation. If $\xi$ is a spatial separation vector, geodesic equation takes the form 
\begin{equation}
\frac{d^{2}\xi^{i}}{dt^{2}}=-{R^i}_{0j0}\xi^{j}.
\label{geodesicdevmemory}
\end{equation}
By substituting Eq.(\ref{Riemannhdg}) into Eq.(\ref{geodesicdevmemory}) and then carrying out the integrals twice reads \cite{Kilicarslan}
\begin{equation}
\begin{aligned}
\Delta\xi^{i}&=\int_{-\infty} ^{U}dU'\int_{-\infty} ^{U'}dU''\frac{d^{2}\xi^{i}}{dU''^{2}}
=\frac{2\pi}{(2\pi r)^{\frac{D-2}{2}}}\bigg(\frac{d^{\frac{D-4}{2}}}{dU^{\frac{D-4}{2}}}- (m_\beta)^{\frac{D-4}{2}}e^{-m_\beta r}\bigg)\Delta_j^i\Theta(U)\xi^{j},
\label{memoryeffect}
\end{aligned}
\end{equation} 
here, $\Delta_j^i$ are spatial components of the memory tensor. Observe that theory gives non-trivial memory effect and memory is reduced by massive spin-$2$ mode compared to GR. In four dimensions, memory takes the form 
\begin{equation}
\begin{aligned}
\Delta\xi^{i}=\frac{1}{r}\bigg(1- e^{-m_\beta r}\bigg)\Delta_j^i\Theta(U)\xi^{j}.
\label{memoryeffect}
\end{aligned}
\end{equation} 
In the large separation limits, the memory reproduces GR result \cite{Garfinkle}, whereas at small distances, it is different. On the other hand, in the $m_\beta\to \infty$ limit, the usual Einsteinian form can be obtained for memory as expected. 

\noindent{\textbf{3. Memory effect in IDG }}

 \noindent We now consider memory effect for particle scattering in IDG as a function of mass scale of non-locality. The Lagrangian density of IDG is \cite{Biswas1}
\begin{equation}
\mathcal{L}= \sqrt{-g}\bigg[\frac{M^2_P}{2}  R\ +\frac{1}{2} R F_1 (\Box) R + \frac{1}{2} R_{ab} F_2(\Box) R^{ab}
       + \frac{1}{2}C_{abcd} F_3(\Box) C^{abcd}+\mathcal{L}_{matter}\bigg],
\end{equation}
where $M_P$ is the Planck mass, $C_{abcd}$ is the Weyl tensor, $R_{ab}$ is the Ricci tensor and $R$ is the scalar curvature. The infinite derivative functions $F_i(\Box)$, which are analytic functions of the d'Alembartian operator, are given as
\begin{equation}
F_i(\Box)=\sum_{n=1}^{\infty}f_{i_n}\frac{\Box^n}{M^{2n}}, 
\label{idfunc}
\end{equation} 
in which $f_{i_n}$ are dimensionless coefficients and $M$ is the mass scale of non-locality. The linearized field equations about a Minkowski background yields \cite{Biswas1}
\begin{equation}
a(\Box)R^L_{ab}-\frac{1}{2}\eta_{ab}c(\Box)R^L-\frac{1}{2}f(\Box)\partial_a\partial_b R^L=\kappa T_{ab},
\label{FeqIDG}
\end{equation}
here non-linear functions are given as
\begin{equation}
\begin{aligned}
 &a(\Box) =1 + M^{-2}_P \left(F_2(\Box) 
        + 2  F_3(\Box)\right) \Box, \\
&        c(\Box) = 1 - M^{-2}_P\left(4 F_1(\Box) +  F_2(\Box)  
        - \frac{2}{3}F_3(\Box)\right)\Box,\\
&        f(\Box) =M^{-2}_P \left(4F_1(\Box) + 2F_2(\Box) +\frac{4}{3} F_3(\Box)\right),
        \end{aligned}
        \label{relations}
\end{equation}
which leads to the constraint $a(\Box)-c(\Box) = f(\Box)\Box$. After substituting the relevant linearized curvature tensors (\ref{lineartensor}) into (\ref{FeqIDG}), linearized field equations can be obtained 
\begin{equation}
\begin{aligned}
  &\frac{1}{2} \bigg[a(\Box)\left(\Box h_{ab} 
        -\partial_d
        \left(\partial_a h^d\,_b + \partial_b h^d\,_a \right)\right)
        + c(\Box) \left(\partial_a \partial_b h + \eta_{ab} \partial_d \partial_e 
        h^{de}-\eta_{ab} \Box h\right)\\&+ f(\Box) \partial_a \partial_b 
 \partial_d \partial_e 
h^{de}\bigg]=-\kappa T_{ab} .
\label{IDGfeq1}
\end{aligned}
\end{equation}
 Note that if we choose $a(\Box)=c(\Box)$, the GR propagator can be recovered in the large separation limit without introducing extra DOF. In the de Donder gauge, the linearized field equations (\ref{IDGfeq1}) can be recast as
\begin{equation}
  a(\Box){\cal{G}}^L_{ab}=\kappa T_{ab}.
  \label{IDGfeq2}
\end{equation}
where $T_{ab}$ is conserved source ( $\partial_a T^ {a b} =0$). Manipulation of the equation (\ref{IDGfeq2}) yields
 \begin{equation}
 a(\Box) \Box h_{ab}=-2\kappa(T_{ab}-\frac{1}{2}\eta_{ab}T) =-16\pi\tilde{T}_{ab},
  \label{IDGfeq3}
\end{equation}
which is the equation that we shall work with. The retarded solution to Eq.(\ref{IDGfeq3}) is
\begin{equation}
h_{ab} = 16\pi\int G_{ab}{}^{cd}(x,x')\tilde{T}_{cd}(x')d^{D}x'.
\end{equation}
Here, $G_{ab}{}^{cd}(x,x')$ is the retarded Green's function of tensorial wave-type equation (\ref{IDGfeq3}) and it is defined as 
\begin{equation}
G_{ab}{}^{cd}(x,x')=\eta_{a}{}^{c}\eta_{b}{}^{d}G(x,x'),
\end{equation}
where $\eta_{a}{}^{c}$ is the parallel propagator. The retarded Green's function of linearized IDG equation is
 \begin{equation}
  G_R(x,x')=\frac{1}{4\pi r }\mbox{erf}(\frac{Mr}{2}) \delta(t-t'-r),
 \end{equation}
 where $\mbox{erf} (r)$ is the error function given by the integral
\begin{equation}
\mbox{erf} (r)=\frac{2}{\sqrt{\pi}}\int_0^r e^{-k^2}dk.
\end{equation}
To calculate the memory effect, let us now again consider the energy momentum tensor given in Eq.(\ref{energy-momentum}). 
Upon using these, the retarded solution for IDG will read
\begin{equation}
  h_{ab}(x)=\frac{4}{r}\bigg(\tilde{\alpha}_{ab}\Theta(U)+\tilde{\beta}_{ab}\Theta(-U)\bigg)\mbox{erf}(\frac{Mr}{2}) ,
  \label{Field1}
 \end{equation}
where we have defined two tensors
\begin{equation}
\begin{aligned}
&\tilde{\alpha}_{ab}(\hat{\mathbf{r}}) \equiv\sum_{(i)out}\frac{d\tau^{(i)}}{dt}\Big(\frac{m^{\rm out}_{(i)}}{1-\hat{\mathbf{r}}\cdot\mathbf{v}^{(i)}}\Big)\bigg(u_{a}^{(i)}u_{b}^{(i)}+\frac{1}{2}\eta_{ab}\bigg),\\&
\tilde{\beta}_{ab}(\hat{\mathbf{r}}) \equiv\sum_{(j)out}\frac{d\tau^{(j)}}{dt}\Big(\frac{m^{\rm out}_{(j)}}{1-\hat{\mathbf{r}}\cdot\mathbf{v}^{(j)}}\Big)\bigg(u_{a}^{(j)}u_{b}^{(j)}+\frac{1}{2}\eta_{ab}\bigg).
\end{aligned}
\end{equation}
 Clearly, at this stage, there is only one difference between the IDG and the usual GR due to the error function in Eq.(\ref{Field1}). In fact, for the large separations, the retarded solution reduces the form of usual GR, but for the small distances the solution converges to a constant. The linearized Riemann tensor for metric perturbation (\ref{Field1}) can be calculated, up to ${\cal{O}}(\frac{1}{r^2})$, as 
\begin{equation}
\begin{aligned}
\partial_d\partial_a\bigg(\frac{\mbox{erf}(\frac{Mr}{2})}{r}\Theta(U)\bigg)=&\bigg(\delta'(U)K_aK_d\frac{\mbox{erf}(\frac{Mr}{2})}{r}-\frac{M}{\sqrt{\pi}r} \delta(U)(K_ar_d+K_dr_a)e^{-\frac{M^2r^2}{4}}\\&-\frac{M^3}{2\sqrt{\pi}} r_ar_d\Theta(U)e^{-\frac{M^2r^2}{4}}\bigg).
\end{aligned}
\end{equation}
Consequently, to leading order, the linearized Riemann tensor of the retarded metric perturbation can be obtained as
\begin{equation}
\begin{aligned}
R_{abcd}=&4K_{[a}\Delta_{b][c}K_{d]}\delta'(U)\frac{\mbox{erf}(\frac{Mr}{2})}{r}-4\bigg(\frac{M}{\sqrt{\pi}\,r} K_{[a}\Delta_{b][c}r_{d]}\delta(U)+\frac{M}{\sqrt{\pi}\,r}K_{[d}\Delta_{b][c}r_{a]}\delta(U)\\&+\frac{M^3}{2\sqrt{\pi}}\,r_{[a}\bar{\alpha}_{b][c}r_{d]}\Theta(U)+\frac{M^3}{2\sqrt{\pi}}\,r_{[a}\bar{\beta}_{b][c}r_{d]}\Theta(-U)\bigg)e^{-\frac{M^2r^2}{4}},
 \label{Riemann}
 \end{aligned}
\end{equation}
where  $\Delta_{ab}$, $\bar{\alpha}_{ab}$ and $\bar{\beta}_{ab}$ are given in equation (\ref{memorytensor}). On the other hand, the relative displacement between two massive test particles at rest is described by geodesic deviation equation. 
By inserting Eq.(\ref{Riemann}) into Eq.(\ref{geodesicdevmemory}) and later integrating this equation twice, one eventually obtains \cite{Kilicarslan:2018yxd}
\begin{equation}
\begin{aligned}
\Delta\xi^{i}&=\int_{-\infty} ^{U}dU'\int_{-\infty} ^{U'}dU''\frac{d^{2}\xi^{i}}{dU''^{2}}
=\frac{1}{r} \mbox{erf}(\frac{Mr}{2})\Delta_j^i\Theta(U)\xi^{j},
\label{memoryeffect}
\end{aligned}
\end{equation} 
where $\Delta_j^i$ are spatial components of the memory tensor. Note that the relative separation of test particles have non-trivial change which is defined by the memory tensor.
The memory is reduced via error function compared to pure GR. In the large separation limits as $r\to \infty$, $\mbox{erf}(r)\to 1$ , the memory takes the usual Einsteinian form as expected. On the other hand, since IDG is a small scale modification of GR,  for lower bound on mass scale of non-locality ($M>4keV$) \cite{Edholm:2018qkc}, the memory reproduces the GR result above at atomic distances.

\noindent{\textbf{{4. Results and conclusions}}

 \noindent The studies on gravitational memory effect have recently taken more attention since there is a hope that it could be measured by advanced LIGO. Here we investigate the memory effect in higher derivative gravity and IDG and give full details of computation whose results were given in  \cite{Kilicarslan, Kilicarslan:2018yxd}. We have computed memory effect in higher derivative gravity and shown that memory is different from pure GR result due to massive spin-$2$ mode whose mass reduces the memory. In the large separations, memory takes the same form as in pure GR. On the other hand, we have demonstrated that gravitational memory in IDG depends on the mass scale of non-locality and hence it is different from GR: memory is modified by error function. The memory returns to usual GR result at sufficiently large separations.

\noindent{\textbf{{Acknowledgements}}

\noindent We would like to thank Bayram Tekin and Suat Dengiz for useful discussions, comments and suggestions.

\end{document}